\documentclass[aps,preprint,nofootinbib,showpacs,superscriptaddress,prc]{revtex4-1}
\date{\today}
\usepackage{amsmath}
\usepackage{amsfonts}
\usepackage{amssymb}

\begin{document}

\title{Alternate proof of the Rowe-Rosensteel proposition and seniority conservation}

\author{Chong Qi}
\email{chongq@kth.se}
\affiliation{KTH (Royal Institute of Technology), Alba Nova University Center,
SE-10691 Stockholm, Sweden}
\affiliation{School of Physics, and State Key Laboratory of Nuclear
Physics and Technology, Peking University, Beijing 100871, China}
\author{X.B. Wang}
\affiliation{School of Physics, and State Key Laboratory of Nuclear
Physics and Technology, Peking University, Beijing 100871, China}
\author{Z.X. Xu}
\affiliation{KTH (Royal Institute of Technology), Alba Nova University Center,
SE-10691 Stockholm, Sweden}
\author{R.J. Liotta}
\affiliation{KTH (Royal Institute of Technology), Alba Nova University Center,
SE-10691 Stockholm, Sweden}
\author{R. Wyss}
\affiliation{KTH (Royal Institute of Technology), Alba Nova University Center,
SE-10691 Stockholm, Sweden}
\author{F.R. Xu}
\affiliation{School of Physics, and State Key Laboratory of Nuclear
Physics and Technology, Peking University, Beijing 100871, China}
\begin{abstract}
For a system with three identical nucleons in a single-$j$ shell, the states can be written as the angular momentum coupling of a nucleon pair and the odd nucleon. The overlaps between these non-orthonormal states form a matrix which coincides with the one derived by Rowe and Rosensteel [Phys. Rev. Lett. {\bf 87}, 172501 (2001)]. The propositions they state are related to the eigenvalue problems of the matrix and dimensions of the associated subspaces. In this work, the propositions will be proven from the symmetric properties of the $6j$ symbols. Algebraic expressions for the dimension of the states, eigenenergies as well as conditions for conservation of seniority can be derived from the matrix.
\end{abstract}

\maketitle

\section{Introduction}

It is well known that seniority remains a good quantum number for systems with identical fermions in a single-$j$ shell when $j\leq7/2$, irrespective of interactions used~\cite{Talmi63,Talmi93}. This property of seniority conservation is no longer valid in shells with $j\geq 9/2$.
In these cases,  the rotationally-invariant interaction has to satisfy a certain number of linear constraints to conserve seniority. The number turns out to be [(2j-3)/6] ($[n]$ denotes the largest integer not exceeding $n$)~\cite{Talmi93}, which is related to the number of states with total spin $I=j$ for three identical nucleons in a single-$j$ shell~\cite{Gino93,Zhao03a,Zam05b}. This indicates that the interactions that mix seniority are only a small fraction of the total two-body interactions~\cite{Talmi93}. 

In past decades, intensive works have been done in deriving algebraic conditions for conservation of seniority~\cite{Talmi93,Isa08,Rowe01,Rowe03,Zhao04}. It is thus found that seniority is still approximately conserved in high-$j$ shells~\cite{Talmi93}.  In $j=9/2$ shell, seniority is exactly conserved in some eigen states even with seniority-nonconserving interactions~\cite{Zam08,Esc06}.  Hence, seniority is still a valuable symmetry in studying medium heavy and heavy semi-magic nuclei. The special advantage of conserving seniority is that in some important cases the states are uniquely defined by seniority and other quantum numbers. This enables us to explore exact analytic expressions for the eigenenergies and wave functions of involved states, which may help in understanding the underlying nuclear structure.
Along this line, works have also been done in studying algebraic expressions for the number of states for identical nucleons in a single-$j$ shell~\cite{Gino93,Zam05b,Zhao05a,Zhao05b,Zhao05c,Zhao03a,Zhao03b,Talmi}.  Ref.~\cite{Talmi10} explored the conditions where the eigenenergy can be expressed as rational functions of two-body interaction.

More specifically, 
in a recent work~\cite{Rowe01,Rowe03}, Rowe and Rosensteel show that the number of linear constraints and algebraic expressions for conservation of seniority can be derived with the quasispin tensor decomposition of the two-body interaction. (Introductions to the quasispin scheme can also be found in Talmi's book~\cite{Talmi93} and recent publications~\cite{Zam05a}.) They proposed a matrix which can project the eigen vectors to two quasispin subspaces and stated that the eigenvalues of the matrix must equal to 2 or -1. In this work, algebraic expressions for the number of states and for conservation of seniority will be explored  from a decomposition of the total angular momentum. A matrix similar to that of Ref.~\cite{Rowe01} can be constructed from the decomposition. The eigenvalue problems of the matrix will be explored in a general way with symmetric properties of angular momentum coupling coefficients. We will also show that  algebraic expansions for state energies and wave functions can be constructed with the help of this matrix.
A short introduction to the framework used in this work will be given first.

\section{Number of states for three fermions in a single shell}

We will focus on systems with three identical fermions.  In constructing the basis states we follow the description of Refs.~\cite{Blom84,Liotta81} where it has been referred to as the multistep shell-model.
For a system with three identical nucleons, the states can be given as the angular momentum coupling (tensor product) of a single nucleon and one nucleon pair as~\cite{Blom84} (see also Ref.~\cite{Zhao04})
\begin{equation}
 | j^2(J_{\alpha})j;I \rangle = \left[a^{\dag}_j\times A^{\dag}_{J_{\alpha}}\right]^I|0\rangle,
\end{equation}
where $a^{\dag}_j$ and $A^{\dag}_{J_{\alpha}}=\frac{\sqrt{2}}{2}[a^{\dag}_j\times a^{\dag}_j]^{J_{\alpha}}$ are nucleon and pair creation operators, respectively, and $I$ is the total spin. $J_{\alpha}$ are even integers which can take values from $0$ to $2j-1$. We have neglected the magnetic quantum numbers for simplicity.  These bases are overcomplete and are not orthonormal to each other.
The overlap between two different bases is given as
\begin{eqnarray}\label{3ov}
\nonumber B^j_I(J_{\alpha}J_{\alpha}')&=&\langle j^2(J_{\alpha})j;I|j^2(J'_{\alpha})j;I\rangle\\
 &=&\delta_{J_{\alpha}J_{\alpha}'}+2M^j_I(J_{\alpha}J_{\alpha}'),
\end{eqnarray}
and
\begin{equation}\label{m2}
M^j_I(J_{\alpha}J_{\alpha}')=\hat{J}_{\alpha}\hat{J}_{\alpha}'
\left\{
\begin{array}{ccc}
j&j&J_{\alpha}'\\
j&I&J_{\alpha}\\
\end{array}
\right\},
\end{equation}
where $\hat{J}=\sqrt{2J+1}$.
The dimensions of the matrix $B$ and $M$ are the same. They equal the number of even integers between $|j-I|$ and $j+I$, i.e., $D=[(3j+1-I)/2]$ and $(2I+1)/2$ for $j\leq I$ and $j\geq I$, respectively. The similarity between overlap matrix elements and coefficients of fractional parentage (i.e., Eq.~(15.10) of Ref.~\cite{Talmi93}) is rather straightforward. 

A matrix similar to Eq.~(\ref{m2}), which was denoted as $M^{\Omega}_{\gamma J}$ and related to the overlap matrix by $M^{\Omega}_{\gamma J} = 2M^j_j(\gamma J)$, has been introduced in Refs.~\cite{Rowe01,Rowe03} with the quasispin tensor decomposition of the two-body interaction. The matrix can project the eigen wave functions into two subspaces with the quasispin of $S=0$ and 2. The first proposition of Refs.~\cite{Rowe01,Rowe03} reads: the eigenvalues of matrix $M^{\Omega}$ must equal to $-1$ or $2$. 
The second proposition is closely related to the first one, which deals with the dimensions of the two subspaces and the number of seniority nonconserving interactions~\cite{Rowe03}. In this work we focus on only these propositions of Refs.~\cite{Rowe01,Rowe03} but we should mention that other results have been discussed by the authors.

We start from the sum rules (orthogonality relations) of $6j$ symbols.
With Eqs.~(10.13) and (10.14) of Ref.~\cite{Talmi93}, we have the following relations
\begin{eqnarray}\label{ms1}
\nonumber \sum_JM^j_I(J_{\alpha}J)M^j_I(JJ_{\alpha}')&=&\hat{J}_{\alpha}\hat{J}_{\alpha}'\sum_J(2J+1)\left\{
\begin{array}{ccc}
j&j&J_{\alpha}\\
j&I&J\\
\end{array}
\right\}\left\{
\begin{array}{ccc}
j&j&J_{\alpha}'\\
j&I&J\\
\end{array}
\right\}\\
&=&\delta_{J_{\alpha}J_{\alpha}'},
\end{eqnarray}
and
\begin{equation}\label{ms2}
\sum_{J, {\rm even}}M^j_I(J_{\alpha}J)M^j_I(JJ_{\alpha}')-\sum_{J, {\rm odd}}M^j_I(J_{\alpha}J)M^j_I(JJ_{\alpha}')=M^j_I(J_{\alpha}J_{\alpha}'),
\end{equation}
where in the first relation the summation is over both even and odd values of spin $J$.
Immediately, we have
\begin{equation}\label{m}
2M^2-M-\mathcal{I}=0,
\end{equation}
where $\mathcal{I}$ denotes a unit matrix.
Above equation indicates that the eigenvalues of the matrix $M$ are equals to $1$ or $-1/2$. This essentially completes our proof of the 
Rowe-Rosensteel proposition. The special feature of our work is that the matrix $M$ is introduced in a different physical framework and the proposition is proven in a more general form. The proposition of Refs.~\cite{Rowe01,Rowe03} corresponds to the special case of $I=j$.

With Eqs.~(\ref{3ov}) and (\ref{m}) , it can be easily recognized that the eigenvalues of the matrix $B$ are equal to $3$ or $0$.
A set of orthonormal bases $|\gamma_n\rangle_I$ can be derived by diagonalizing the overlap matrix $B$. We have
\begin{equation}
B|\gamma_n\rangle_I=3|\gamma_n\rangle_I.
\end{equation}
The dimension of the orthonormal bases will be denoted as $D_I(j)$.
It equals to the rank of overlap matrix $B$.
We have
\begin{equation}
D_I(j)={\rm rank}(B^j_I)=\frac{1}{3}{\rm tr}(B^j_I).
\end{equation}
where ${\rm tr}(B)$ denotes the trace of the corresponding matrix. We note that a similar relation has been derived in Ref.~\cite{Zam05c} by employing a special spin-independent two-body interaction.

Based on symbolic calculations with Racah's analytic formula for the $6j$ symbol~\cite{Racah} (Eq.~(10.26) in Ref.~\cite{Talmi93}), we found that the following relations hold.
For $I\leq j$, we have
\begin{equation}
{\rm tr}(M)=\sum_{J_{\alpha}, {\rm even}}(2J_{\alpha}+1)
\left\{
\begin{array}{ccc}
j&j&J_{\alpha}\\
j&I&J_{\alpha}\\
\end{array}
\right\}=
\left\{
\begin{array}{rcc}
-1/2&~&{\rm mod}(2I,6)=1,\\
1/2& &{\rm mod}(2I,6)=3,\\
0&&{\rm mod}(2I,6)=5.\\
\end{array}
\right.
\end{equation}
A similar sum rule has been found for the special case of $I=j$ in Ref.~\cite{Zhao03b} (see also Ref.~\cite{Zhao04}).
For states with $I\leq j$, the corresponding dimensions can be written as
\begin{equation}\label{d1}
D_I(j)=\frac{2I+1+4{\rm tr}(M)}{6}=
\left\{
\begin{array}{lcc}
\left[I/3\right]&~&{\rm mod}(2I,6)=1\\
\left[I/3\right]+1& &{\rm mod}(2I,6)=3\\
\left[I/3\right]+1&&{\rm mod}(2I,6)=5\\
\end{array}
\right\}=\left[\frac{I+2}{3}\right]=\left[\frac{2I+3}{6}\right].
\end{equation}
For $I\geq j$, we have
\begin{equation}
{\rm tr}(M)=\sum_{J_{\alpha}, {\rm even}}(2J_{\alpha}+1)
\left\{
\begin{array}{ccc}
j&j&J_{\alpha}\\
j&I&J_{\alpha}\\
\end{array}
\right\}=
\left\{
\begin{array}{rcl}
1/2&~&{\rm mod}(3j-3-I,6)=0\\
-1& &{\rm mod}(3j-3-I,6)=1\\
0&&{\rm mod}(3j-3-I,6)=2~ {\rm or}~3\\
-1/2&&{\rm mod}(3j-3-I,6)=4~ {\rm or}~5.\\
\end{array}
\right.
\end{equation}
Similarly to Eq.~(\ref{d1}), we have
\begin{eqnarray}
\nonumber D_I(j)&=&\frac{[(3j+1-I)/2]+2{\rm tr}(M)}{3}\\
\nonumber &=&
\left\{
\begin{array}{lcl}
\left[\frac{3j+3-I}{6}\right]&~&{\rm mod}(3j-3-I,6)=0\\
\left[\frac{3j+3-I}{6}\right]-1& &{\rm mod}(3j-3-I,6)=1\\
\left[\frac{3j+3-I}{6}\right]&&{\rm mod}(3j-3-I,6)=2~ {\rm or}~3\\
\left[\frac{3j+3-I}{6}\right]&&{\rm mod}(3j-3-I,6)=4~ {\rm or}~5\\
\end{array}
\right.\\
&=&\left[\frac{3j+3-I}{6}\right]-\delta_{{\rm mod}(3j-3-I,6),1}.
\end{eqnarray}
Above algebraic expressions for state dimensions were first empirically obtained in Ref.~\cite{Zhao03a} and analytically proved in Ref.~\cite{Talmi}.

\section{Expression for the eigenenergies}

The Hamiltonian matrix can be written as
\begin{equation}
\langle\gamma_m|\hat{V}|\gamma_n\rangle_I = \langle\gamma_m| j^2(J_{\alpha})j;I \rangle \langle j^2(J_{\alpha})j;I |\hat{V}| j^2(J_{\alpha}')j;I \rangle B^j_I(J_{\alpha}'J_{\alpha}'') \langle j^2(J_{\alpha}'')j;I|\gamma_n\rangle_I,
\end{equation}
where
\begin{equation}\label{h2}
\langle j^2(J_{\alpha})j;I |\hat{V}| j^2(J_{\alpha}')j;I \rangle = V_{J_{\alpha}'}B^j_I(J_{\alpha}J_{\alpha}').
\end{equation}
$V_J=\langle j^2;J|\hat{V}|j^2;J\rangle$ are two-body matrix elements with $|j^2;J\rangle$ denoting coupled two-body states. Although the matrix defined by Eq.~(\ref{h2}) is not symmetric, one can easily recognize that the Hamiltonian matrix is symmetric since the overlap $B^j_I$ is a symmetric matrix.

For a system with three identical fermions in a single-$j$ shell, we have one $I=j$ state seniority $\nu=1$ and $[(j-1)/3]$ states with seniority $\nu=3$. The conservation of seniority implies that $\langle\gamma_{\nu=1}|\hat{V}|\gamma_{\nu=3}\rangle_j$=0~\cite{Talmi93}.
Based on the overlap matrix, the conservation condition can be written as~\cite{Qi10}
\begin{equation}\label{lc}
\sum_{J>0}B^j_j(0J_{\alpha})\left(B^j_j(0J_{\alpha})-\frac{B^j_j(00)}{B^j_j(0\lambda)}B^j_j(J_{\alpha}\lambda)\right)V_{J_{\alpha}}=0,
\end{equation}
where $\lambda$ denotes a even integer which can take values between $2$ and $(2j-1)$.
For interaction $V_0$, the corresponding coefficient is zero. The total number of linear constraints is calculated to be $n=D_j(j)-1=[(j-1)/3]=[(2j-3)/6]$.
With Eq.~(\ref{3ov}), we found that the expansion coefficients of Eq.~(\ref{lc}) can be written as
\begin{eqnarray}
\nonumber  a_{\lambda J}&=&B^j_j(0J_{\alpha})\left(B^j_j(0J_{\alpha})-\frac{B^j_j(00)}{B^j_j(0\lambda)}B^j_j(J_{\alpha}\lambda)\right)\\
&=&C\left[\frac{4(2J_{\alpha}+1)}{(2j+1)(2j-1)} - \delta_{J_{\alpha}\lambda}-2(2J_{\alpha}+1)
\left\{
\begin{array}{ccc}
j&j&J_{\alpha}\\
j&j&\lambda\\
\end{array}
\right\}
\right]\hat{\lambda},
\end{eqnarray}
where $C$ denotes a constant. A similar analytic expression has been derived in Ref.~\cite{Isa08}. It can be easily seen from the above equation seen that the summation of terms in the bracket is a rational number. As a result, the seniority conservation condition can be expressed as rational functions of the interaction.  The algebraic conservation conditions for $9/2\leq j\leq 13/2$ have been given explicitly in Ref.~\cite{Talmi93,Rowe01}.
The conservation condition for $j=15/2$ can be given as
\begin{equation}
1330V_2-2835V_4-1807V_6+612V_8+3150V_{10}+3175V_{12}-3625V_{14} = 0,
\end{equation}
and
\begin{equation}
77805V_2-169470V_4-85527V_6-4743V_8+222768V_{10}+168025V_{12}-208858V_{14} = 0.
\end{equation}
These two expressions are derived by assuming $\lambda=2$ and $4$, respectively. The conditions can also be expressed as linear combinations of above  expressions and those calculated with other $\lambda$ values.

It can be easily seen that the following sum rules of $6j$ symbols should hold,
\begin{equation}
\sum_{J_{\alpha}\neq0,~ {\rm even}}(2J_{\alpha}+1)
\left\{
\begin{array}{ccc}
j&j&J_{\alpha}\\
j&j&\lambda\\
\end{array}
\right\} = \frac{1}{2j+1}+\frac{1}{2},
\end{equation}
and
\begin{equation}
\sum_{J_{\alpha},~{\rm even}}(2J_{\alpha}+1)
\left\{
\begin{array}{ccc}
j&j&J_{\alpha}\\
j&j&\lambda\\
\end{array}
\right\} = \frac{1}{2}.
\end{equation}
These relation are derived with Racah's algebraic expression of $6j$ symbols. But they can also be derived from Eqs. (\ref{ms1}) and (\ref{ms2}) of this work as well as from Eqs.~(10.13) and (10.14) (or Eqs.~(10.20) and (10.21)) of Ref.~\cite{Talmi93}.
With above relations we have
\begin{equation}
\sum_{J_{\alpha}} a_{\lambda J_{\alpha}} = 0.
\end{equation}

The eigenenergies and eigen vectors can be calculated by diagonalizing the Hamiltonian matrix.
In a single-$j$ shell, the energy of a state with total angular momentum $I$  can be written as~\cite{Talmi93,Talmi10,Zhao04,Qi10}
\begin{equation}
E_I=C_{J_{\alpha}}^IV_{J_{\alpha}},
\end{equation}
where $C_{J_{\alpha}}$ are expansion coefficients which are usually irrational functions of the two-body interactions~\cite{Qi10,Talmi10}. 
The projection of the eigen wave function $\Psi$ onto the basis states is given as
\begin{equation}\label{f}
\langle\Psi_n|j^2(J_{\alpha})j;I\rangle = F^j_I(J_{\alpha}) = \langle\Psi_n|\gamma_m\rangle\langle\gamma_m |j^2(J_{\alpha}')j;I\rangle B^j_I(J_{\alpha}'J_{\alpha}).
\end{equation}
The expansion coefficient $C$ is related to the projection $F$ by
\begin{equation}
C_{J_{\alpha}} =|F^j_I(J_{\alpha})|^2.
\end{equation}
We have
\begin{equation}
\sum_{J_{\alpha}}C_{J_{\alpha}}=\sum_{J_{\alpha}}|F^j_I(J_{\alpha})|^2=3.
\end{equation}
For a state with dimension $D_I(j)=1$, it can be easily understood from Eq.~(\ref{f}) that the corresponding energy expression would be uniquely defined with expansion coefficients $C$ being related to the overlap matrix elements.

If seniority was conserved in the interaction, the system would be exactly solvable. These have been intensively studied in textbooks~\cite{Talmi93} and recent publications~\cite{Zhao04} and will not be detailed here. As an illustration, 
the wave function of unique $\nu=1$, $I=j$ state for three-identical-nucleon system in any single-$j$ shell satisfies the relation
\begin{equation}
\langle\Psi_{\nu=1}|j^2(J_{\alpha})j;j\rangle=\frac{B^j_j(0J_{\alpha})}{\sqrt{B_j(00)}} = \sqrt{\frac{2j+1}{2j-1}}\left(\delta_{J_{\alpha,0}} - \frac{2\hat{J}_{\alpha}}{2j+1}\right).
\end{equation}
Correspondingly, the energy expression can be given as
\begin{equation}
E = \frac{2j-1}{2j+1}V_0+\sum_{J>0}\frac{4(2J+1)}{(2j+1)(2j-1)}V_J.
\end{equation}

\section{Summary}

In summary, algebraic expressions for the number of states,  eigenenergies and condition for conservation of seniority have been explored  for  systems with three identical nucleons in a single-$j$ shell from a decomposition of the total angular momentum. A matrix similar to that of Ref.~\cite{Rowe01} have been constructed from the decomposition.  The eigenvalue problem of the matrix was explored in a general way with symmetric properties of angular momentum coupling coefficients.

\section*{ACKNOWLEDGMENTS}
This work has been supported by the Swedish Research Council (VR), the Chinese Major State Basic
Research Development Program under Grant 2007CB815000 and the
National Natural Science Foundation of China under Grant Nos.
10735010 and 10975006. C.Q. thanks Prof. Yu-Min Zhao (Shanghai) for valuable comments.

\end{document}